\documentclass[10pt,conference]{IEEEtran}
\usepackage{ifthen}
\usepackage[cmex10]{amsmath} % Use the [cmex10] option to ensure compliance with IEEE Xplore (see bare_conf.tex)
\usepackage{mypackages}
\usepackage{mytheorem}
\usepackage{mycommands}

\renewcommand{\baselinestretch}{0.97}

\newcommand{\figwidth}{0.844\columnwidth}

\begin{document}

\title{Adaptive Decoding of LDPC Codes \\ with Binary Messages}

\author{
\IEEEauthorblockN{Ingmar Land, Gottfried Lechner}
\IEEEauthorblockA{
  Institute for Telecommunications Research \\
  University of South Australia \\
  Adelaide, Australia \\
  Email: \{ingmar.land,gottfried.lechner\}@unisa.edu.au}
\and
\IEEEauthorblockN{Lars Rasmussen}
\IEEEauthorblockA{
  KTH, Royal Institute of Technology \\
  School of Electrical Engineering \\
  Stockholm, Sweden \\
  Email: lars.rasmussen@ee.kth.se}
}

\maketitle

\begin{abstract}
A novel adaptive binary decoding algorithm for LDPC codes is proposed, which reduces the decoding complexity while having a comparable or even better performance than corresponding non-adaptive alternatives.  In each iteration the variable node decoders use the binary check node decoders multiple times; each single use is referred to as a sub-iteration.  To process the sequences of binary messages in each iteration, the variable node decoders employ pre-computed look-up tables.  These look-up tables as well as the number of sub-iterations per iteration are dynamically adapted during the decoding process based on the decoder state, represented by the mutual information between the current messages and the syndrome bits.  The look-up tables and the number of sub-iterations per iteration are determined and optimized using density evolution.  The performance and the complexity of the proposed adaptive decoding algorithm is exemplified by simulations.   
\end{abstract}

%---------------------------------------------------------------
\section{Introduction}

Low-density parity-check (LDPC) codes together with message passing decoders are among the most powerful channel coding schemes known \cite{Gallager:1962,Davey/MacKay:1998}.  They allow performance close to the Shannon capacity while the decoding complexity grows only linearly in the code length.  Even though the decoding complexity of LDPC codes is already relatively low, it may still be too high for some high speed applications, e.g. fiber-optics.  A further reduction of the decoding complexity leads usually to performance losses.  Among the class of message-passing algorithms, the sum-product algorithm achieves the best performance but requires real-valued messages.  In contrast, bit-flipping algorithms pass only single-bit messages but also suffer poorer performance \cite{Ardakani2005,Lechner/Pedersen/Kramer:ISIT:2007}.  

In \cite{Lechner/Land/Rasmussen:ISTC:2008} we introduced a new type of binary message passing algorithm.  The variable node decoders send binary messages to the check node decoders, which perform simple modulo-2 addition.  In each iteration, however, the variable node decoders may use the check node decoders multiple times, where each single use is referred to as one \emph{sub-iteration}.  The binary messages per sub-iteration are generated similarly to ideas in stochastic decoding \cite{SharifiTehrani:2006p716}.  In addition to that, the variable node decoders use pre-computed look-up tables to optimally process the bit messages in each iteration.  All more complex computations are performed in the variable node decoders, which makes this decoding structure attractive for efficient high-speed implementations.

For the analysis of this algorithm it is convenient to collect (for each edge) all binary messages of an iteration into a binary vector message.  Therefore the algorithm is referred to as the \emph{binary vector message passing algorithm} (BVMPA) \cite{Lechner/Land/Rasmussen:ISTC:2008}.  The selection of the number of sub-iterations per iteration (which is equivalent to length of the binary vector messages) allows an easy trade-off between performance and complexity. In \cite{Lechner/Land/Rasmussen:ISTC:2008} the BVMPA is investigated for a fixed length of the binary vector messages.

In the present paper we propose an extension of the BVMPA that reduces the complexity while keeping the same performance or even achieving better performance as compared to the original BVMPA.   Both the look-up table and the number of sub-iterations per iteration are dynamically adapted to the current status of the decoding process, referred to as decoder state.  The key contributions of the present paper to reduce the decoding complexity are:
\begin{enumerate}
\item We define the measure \emph{syndrome information} that is used to track the quality of the current message during the decoding process.  This measure is considered as a decoder state with respect to the decoding process.
\item We determine optimized \emph{look-up tables} to be used when the decoder is in a certain state. 
\item We optimize the \emph{number of sub-iterations per iteration (vector lengths)} to be used when the decoder is in a certain state. 
\end{enumerate}
The optimum look-up tables and the optimum number of sub-iterations are determined using density evolution.  Simulations show that bit error rates of $10^{-6}$ can be achieved with an approximate average of only $50$ sub-iterations, i.e., the decoder converges with only $50$ bits exchanged per edge.

The concept of adapting the decoding process dynamically was also studied in \cite{Yue/Wang:ISTC:2008} for a bit-flipping algorithm.  The authors estimated an extrinsic error probability, and used it to dynamically adjust the bit-flipping threshold in the variable node decoders.  In  \cite{Ardakani/Kschischang:TCOM:2006} the concept of ``gear-shift decoding'' was introduced where the decoder can select among a small set of available decoding algorithms (sum-product algorithm, min-sum algorithm, and bit-flipping algorithm) to reduce the decoding complexity. Both concepts will be incorporated in the BVMPA.

The outline of the paper is as follows.  In Section~\ref{sec:system-model} we introduce the system model and review the decoder structure.  In Section~\ref{sec:decoder-state-decoding} we introduce the concept of decoder states to track the progress of the decoding, and we explain how this can be used directly to improve the actual decoding.  In Section~\ref{sec:message-length-optimization} we address decoding with variable vector lengths, and optimize the vector lengths with respect to the decoder state.  Simulation results for the resulting decoder structure are presented in Section~\ref{sec:simulation-results}.

Throughout the paper, random variables are denoted by uppercase letters, and their realizations are denoted by lowercase letters.   The indices $v$, $c$, $a$, and $e$ stand for variable-node decoder, check-node decoder, a-priori and extrinsic, respectively.  We assume that the reader is familiar with standard knowledge about LDPC codes, see e.g. \cite{Richardson/Urbanke:2008}.

%---------------------------------------------------------------
\section{System Model and Decoder Structure}
\label{sec:system-model}

\subsection{System Model}

We consider regular $(d_v,d_c)$ LDPC codes of code length $N$, where $d_v$ denotes the variable node degree and $d_c$ denotes the check node degree.  (The concepts can be extended to irregular LDPC codes).  The codewords are assumed to be equiprobable and are transmitted over a memoryless symmetric communication channel, e.g. an AWGN channel.  

Denote a code bit as $X \in \{0,1\}$, and the corresponding channel output as $y$.  The L-values of the code bits given the direct observation are
\begin{equation*}
  l_{ch} := L(X|y) = \ln \frac{\Pr(X=0|y)}{\Pr(X=1|y)} .
\end{equation*}
They are called \emph{channel L-values}, and are provided as input to the LDPC decoder.  

%---------------------------------------------------------------
\subsection{Decoder Structure}

For decoding we employ the BVMPA introduced in \cite{Lechner/Land/Rasmussen:ISTC:2008}.  In the following this algorithm is briefly reviewed, and the amendments proposed in the present paper are outlined.  For details on the original BVMPA we refer the reader to \cite{Lechner/Land/Rasmussen:ISTC:2008}.

The iterative decoder for the LDPC code operates on the factor graph of the code defined by a parity-check matrix $H$.  The decoder iterates until the codeword estimate $\hat{\word{x}}$ is an actual codeword, i.e., until $H \hat{\word{x}} = 0$.

The variable node decoders (VNDs) and the check node decoders (CNDs) exchange messages that are binary vectors $\vect{b} \in \{0,1\}^Q$ of length $Q$.  By definition, only the Hamming weight of the vector, $w = w_H( \vect{b} )$, and not the positions of the  ones within the vector, conveys information. Therefore the weight represents, in some sense, the probability that the corresponding code bit has the value one.

The \emph{VND} for a code bit $X$ gets the channel L-value $l_{ch}$ and $d_v$ binary vector messages $\vect{b}_{av,j}$, $j=1,...,d_v$ from the CND.  Each vector $\vect{b}_{av,j}$ with weight $w_{av,j}$ is first converted into an L-value
\begin{align}
  l_{av,j} 
  &:= L( X | \word{b}_{av,j} )
  = L( X | w_{av,j} )                  \notag\\
  &= \ln \frac{ p( w_{av,j} | X=0 ) }
              { p( w_{av,j} | X=1 ) }
  = \ln \frac{ p( w_{av,j} | X=0 ) }
              { p( Q-w_{av,j} | X=0 ) } \notag \\
  &=: f_{W2L} \bigl( w_{av,j} , p( w_{av,j} | X=0 ) \bigr) .
  \label{eq:w2l}
\end{align}
For this conversion, we first use that only the vector weight is relevant, and then that the probability $p(w_{av} | X=0 )$ fulfils a certain symmetry property \cite{Lechner/Land/Rasmussen:ISTC:2008}.  In the following we refer to this mapping from weights to L-values as the \emph{W2L mapping} $f_{W2L}( w_{av} , p( w_{av} | X=0 ) )$, where $w_{av}$ is the weight of the vector message and $p( w_{av} | X=0 )$ is the conditional probability distribution of this weight given $X=0$.  Notice that this distribution changes over the decoder iterations.

Using density evolution, the probability distributions $p( w_{av} | X=0 )$ for each iteration $i$ can be pre-computed, and so can the W2L mappings.  In this way we obtain iteration-dependent W2L mappings $f^{it}_{W2L}( w_{av} , i )$, as applied in \cite{Lechner/Land/Rasmussen:ISTC:2008}.  (The index $it$ stands for ``iteration dependent''.)  Notice that the actual decoding process may not exactly follow the one predicted by density evolution, particularly for medium to short block lenghts, and thus the applied W2L mappings may not be optimal.  This issue will be addressed in Section~\ref{sec:decoder-state-decoding}.

The VND then adds these a-priori L-values to obtain the extrinsic L-value
\begin{equation}
  \label{eq:L-add}
  l_{ev,k} = l_{ch} + \sum_{j=1 , j \ne k}^{d_v} l_{av,j} .
\end{equation}
This L-value is converted into the corresponding probability that the code bit has value one,
\begin{equation}
  \label{eq:l2p}
  p_{ev,k} := \Pr( X=1 | l_{ev,k} )
  = \bigl( 1 + e^{l_{ev,k}} \bigr)^{-1} .
\end{equation}
The desired weight of the extrinsic message is obtained by
\begin{equation}
  \label{eq:p2w}
  w_{ev,k} := \rnd( p_{ev,k} Q ) ,
\end{equation}
where the function $\rnd(\cdot)$ denotes rounding to integers.  The actual vector $\vect{b}_{ev,k}$ is then a random binary vector of length $Q$ and weight $w_{ev,k}$ \cite{Lechner/Land/Rasmussen:ISTC:2008}.  This random element in our decoding algorithm shows similarities to stochastic decoding \cite{SharifiTehrani:2006p716}. 

The \emph{CND} gets $d_c$ binary vector messages $\vect{b}_{ac,j}$, $j=1,...,d_c$.  The extrinsic messages are computed by a simple modulo-2 addition,
\begin{equation}
  \label{eq:cnd-operation}
  \word{b}_{ec,k} =
  \sum_{j=1,j \ne k}^{d_c}\hspace{-4.8ex}\oplus\hspace{+1.5ex} \word{b}_{ac,j} .
\end{equation}
In expectation this operation is optimal \cite{SharifiTehrani:2006p716}.  Notice that this vector addition is a bit-wise operation.  For an efficient implementation, the CND may be a simple bit addition, like in binary message passing decoders, and the VND may use each corresponding CND $Q$ times.  In the following we refer to one of this binary operations as one \emph{sub-iteration}.  One iteration with message length $Q$ thus corresponds to $Q$ sub-iterations.

The complexity of the BVMPA is proportional to the message length times the number of iterations, or equivalently, to the number of sub-iterations until convergence.  Decoding of regular LDPC codes is only critical in the middle of the decoding process where the EXIT curves of the VNDs and the CNDs are close to each other. Correspondingly large vector lengths $Q$ are usually not required at the beginning and at the end of decoding.  By reducing the lengths of the messages when possible without loss in performance, the average number of sub-iterations and thus the average decoding complexity is expected to be reduced.  The optimization of the message lengths and their adaptation during the decoding process is addressed in Section~\ref{sec:message-length-optimization}.

%---------------------------------------------------------------
\section{Decoder-state Dependent Decoding}
\label{sec:decoder-state-decoding}

In the considered decoder structure, the W2L mapping from the weight of the vector message to the L-value at the VND input is not constant but changes over the iterations.  In \cite{Lechner/Land/Rasmussen:ISTC:2008} the iteration number was used to parameterize this mapping, as given in \eqref{eq:w2l}.  The problem with this approach, however, is that an actual instance of the decoding process may not be accurately predicted by density evolution, particularly for medium or short block lengths. Thus the mapping used at a certain iteration number is likely to be suboptimal.  

Rather than using the iteration number to select the W2L mapping, we propose to use a measure that reflects the quality of the messages currently exchanged between the VNDs and the CNDs.  Using this approach, the decoder selects the W2L mapping based on the current message quality, and thus adapts dynamically.

%---------------------------------------------------------------
\subsection{Decoder State Based on Syndrome}

The decoder state has to be a value that can be measured in a real decoder without knowledge of the data transmitted.  In \cite{Yue/Wang:ISTC:2008} the extrinsic error probability was used in a similar context.  We propose to measure information about syndrome bits that results from the assumption that the actual syndrome is not known.  (Of course, the syndrome $H \word{x}$ is known to be zero for every codeword $\word{x}$.)  We will first explain this idea for the case where the messages are L-values, and then for the given case where the messages are binary vectors.

Consider a check node of degree $d_c$ corresponding to the code bits $X_n$, $n=1,2,\ldots,d_c$.    Define the syndrome bit
\begin{equation}
  S := X_1 \oplus X_2 \oplus \cdots \oplus X_{d_c}  ,
  \label{equ:syndrome-bit}
\end{equation}
where by definition of the code, we know that $S=0$.  Consider now the decoder for this check node.  For the time being, assume that this CND obtains the a-priori L-values $l_{ac,n}$, $n=1,2,\ldots,d_c$, for these code bits. (We will later come back to binary vector messages.)  The extrinsic L-value for $X_{d_c}$ is computed as \cite{Hagenauer/Offer/Papke:1996}
\begin{equation}
  l_{ac,d_c} := l_{ac,1} \boxplus l_{ac,2} \boxplus \cdots \boxplus l_{ac,d_c-1} .
  \label{equ:extrinsic-Lvalue}
\end{equation}
Define now
\begin{equation}
  l_s := l_{ac,1} \boxplus l_{ac,2} \boxplus \cdots \boxplus l_{ac,d_c} .
  \label{equ:syndrome-Lvalue}
\end{equation}
This L-value is obvioulsy the extrinsic L-value for the syndrome bit $S$, as defined in \eqref{equ:syndrome-bit}.  It corresponds to the probability of $S$ being zero or one given the a-priori L-values for the $d_c$ code bits (and excluding the knowledge that $S$ is zero by definition of the code).  In this setting with L-values, the probability for $S$ being one can be used to define the decoder state.

Consider now the case that the CND obtains the a-priori binary-vector messages $\vect{b}_{ac,n}$, $n=1,2,\ldots,d_c$, as in the algorithm employed in the present paper.  The extrinsic message for $X_{d_c}$ is computed as
\begin{equation}
  \vect{b}_{ac,d_c} 
  := \vect{b}_{ac,1} \oplus \vect{b}_{ac,2} \oplus \cdots 
     \oplus \vect{b}_{ac,d_c-1} .
  \label{equ:extrinsic-message}
\end{equation}
Similar to the case above with L-values, define now
\begin{equation}
  \vect{b}_s 
  := \vect{b}_{ac,1} \oplus \vect{b}_{ac,2} \oplus \cdots 
     \oplus \vect{b}_{ac,d_c} .
  \label{equ:syndrome-message}
\end{equation}
This message is the extrinsic message for the syndrome bit $S$ given the a-priori messages for the code bits (and excluding the knowledge that $S$ is zero by definition of the code).

These syndrome bit messages are now computed for all check nodes.  By definition of the binary vector messages, only the weight of the vector conveys information but not the positions of the ones.  Therefore we compute the weight of each vector, consider this weight as a random variable $W_s$, and determine the empirical \emph{weight distribution of the syndrome bit messages}, denoted by $\hat{p}_{W_s}(w)$.  As we know that the actual values of the syndrome bits are zero, this is in fact the conditional weight distribution given $S=0$.  This can now be used to define a decoder state.  Notice that this can be computed without knowledge of the transmitted data, as desired.

The decoder state should be a scalar value, for convenience, and it should capture the information in the distribution that is relevant for the iterative decoding process.  Motivated by that, we define the decoder state as
\begin{equation}
  \label{equ:syndrome-information}
  I_S 
  := H \bigl( \tfrac{1}{2}( \hat{p}_{W_s}(w) + \hat{p}_{W_s}(Q-w) ) \bigr)
     - H \bigl( \hat{p}_{W_s}(w) \bigr) ,
\end{equation}
where $H( p )$ denotes the entropy of the probability mass function $p$.  The value $I_S$ is the average bit-wise mutual information between the syndrome bit and the weight of the syndrome message, under the assumption that the codewords are randomly chosen over all binary vectors.  Given this interpretation, we refer to the decoder state defined in \eqref{equ:syndrome-information} as \emph{syndrome information}.

%---------------------------------------------------------------
\subsection{Use of Decoder State in Decoding}

As outlined at the beginning of this section, the W2L mapping 
can now be made decoder-state dependent instead of iteration dependent.  We use density evolution for the BVMPA, see \cite{Lechner/Land/Rasmussen:ISTC:2008}, and extend it such that also the syndrome information is computed.  (As this extension is straightforward, we omit the details here.)  

Using this extension, the W2L mapping can be determined for each syndrome information, where the values of the syndrome information $I_S$ are uniformly quantized for practical reasons.  We denote these mappings by $f^{ds}_{W2L}(w,I_S)$, where $w$ denotes the weight of the extrinsic binary vector message obtained by the VND, and $I_S$ is the current value of the syndrome information.  (The index $ds$ stands for ``decoder state''.)  

These mappings are given to the LDPC decoder as a pre-computed look-up table.  During the decoding process, the syndrome information is computed in each iteration, and the corresponding W2L mapping is applied.  The computational overhead to compute the syndrome information is negligible as this has to be done only once per iteration (once per $Q$ sub-iterations).  In addition to that, the value of $Q$ is relatively small in the relevant cases, and so the look-up tables are relatively small as well. 

Besides using the syndrome information to allow the decoder to select the best mapping, we also use this quantity as an additional stopping criterion, i.e. the decoder stops iterating and declares an error if the syndrome information is not improved within a given number of iterations.

The effect of this approach is exemplified with simulation results for a $(3,6)$ LDPC code of length $N=1000$ for message lengths $Q=10$.  We compare decoding with iteration dependent W2L mappings \cite{Lechner/Land/Rasmussen:ISTC:2008} to decoding with decoder-state dependent mappings.  For the iteration-dependent case, the look-up table comprises one row per iteration ($100$ in our example).  For the decoder-state dependent case, however, the look-up table comprises only one row per quantized syndrome information, where $20$ quantization levels turned out to be sufficient.  Regarding the stopping criterion, the LDPC decoder terminates if the syndrome information has not increased over the previous $10$ iterations.

\begin{figure}[t]
\centering
\psfrag{xlabel}[tc]{${E_{b}}/{N_{0}}$ [dB]}
\psfrag{ylabel}[cb]{bit error rate}
\includegraphics[width=\figwidth]{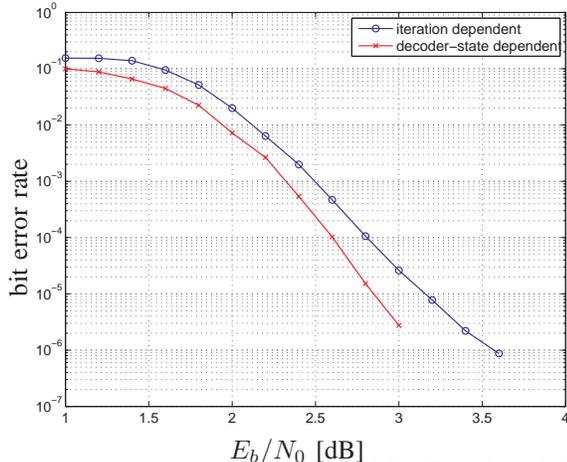}
\caption{Bit error rate of a $(3,6)$ LDPC code of length $N=1000$ for message lengths $Q=10$ with iteration dependent W2L mappings and with decoder-state dependent W2L mappings.}
\label{fig:decstate-W2L-BER}
\end{figure}

\begin{figure}[t]
\centering
\psfrag{xlabel}[tc]{${E_{b}}/{N_{0}}$ [dB]}
\psfrag{ylabel}[cb]{sub-iterations}
\includegraphics[width= \figwidth]{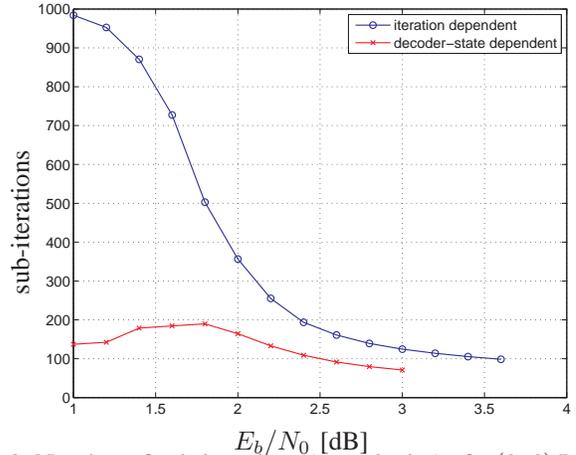}
\caption{Number of sub-iterations (complexity) of a $(3,6)$ LDPC code of length $N=1000$ for message lengths $Q=10$ with iteration dependent W2L mappings and with decoder-state dependent W2L mappings.}
\label{fig:decstate-W2L-NoIt}
\end{figure}

Fig.~\ref{fig:decstate-W2L-BER} shows the resulting bit error rates.  
The decoder-state dependent approach leads to a better performance over the whole SNR range. 
The average number of sub-iterations is depicted in Fig.~\ref{fig:decstate-W2L-NoIt}.
Here, the improvement of the decoder-state dependent approach is even more significant. For low SNR, the additional stopping criterion reduces the average number of iterations while for high SNR, the decoder-state dependent selection of the mapping leads to a faster convergence.

%---------------------------------------------------------------
\section{Message Length Optimization}
\label{sec:message-length-optimization}

In the previous section we reduced the average number of sub-iterations and thus the decoding complexity by employing decoder-state dependent W2L mappings.  In this section we further reduce the decoding complexity by using the smallest message lengths possible without deteriorating performance as compared to the BVMPA with corresponding fixed message length.

The motivation is similar to the one in the previous section.  As previously discussed, only ``the middle part'' of the decoding process is critical for convergence and thus requires messages of higher precision, whereas shorter messages are typically sufficient in the beginning and the end of the decoding process.  

In the following we optimize the message lengths using density evolution.  The LDPC decoder then selects the required message length during the decoding process based on the decoder state.

%---------------------------------------------------------------
\subsection{Method}

Consider the decoding structure described in Section~\ref{sec:system-model} but with an additional degree of freedom; the length $Q$ of the binary vector messages is not fixed but may change over iterations.  Denote $Q_i$ the message length used at iteration $i$, and denote $Q_{\max}$ the maximum message length.

The \emph{optimization problem} is then: determine the sequence of message lengths $\vect{Q} = [ Q_1 , Q_2 , Q_3 , \ldots ]$ such that the overall number of sub-iterations, $\sum_j Q_j$, is minimized for successful decoding ($I_S=1$).  

A complete search is obviously not feasible and thus, simplifying assumptions are required.  In the following we develop a trellis-based method similar to \cite{Ardakani/Kschischang:TCOM:2006} to solve this problem.  First we define the \emph{cost function}
\begin{equation}
  \label{equ:cost-function}
  c_i = \sum_{j=1}^{i} Q_j ,
\end{equation}
for the path $\vect{Q}_i = [Q_1 , Q_2 , \ldots , Q_i ]$ of length $i$; i.e. a path corresponding to $i$ iterations and a total of $c_i$ sub-iterations.  Consider now the tree representing all possible sequences $\vect{Q}$.  Assume that the tree has been completely searched up to a certain depth $i$.  Then at this depth there may be paths that have the same syndrome information but different costs.  In the EXIT chart method it is usually assumed that the mutual information represents the most important parameter of a probability distribution.  Following this idea, among the paths that have the same syndrome information, we keep only the one with the least cost, and discard the other ones.

As the syndrome information is a real value, it is unlikely that many paths will have exactly the same value.  To further reduce the search complexity, we quantize the syndrome information, and refine the above selection criterion. Among all paths (at the same depth) that have the same quantized syndrome information, we only keep the one with the least cost.

Using these simplifications, it is straightforward to solve the optimization problem with a trellis-based search, where the trellis states are the quantized values of the syndrome information.  With a sufficient large number of quantization levels, the resulting path $\vect{Q}^*$ can be expected to be very close to the optimum one.  For convenience, we simply refer to $\vect{Q}^*$ as the optimum path.

%---------------------------------------------------------------
\subsection{Application in Decoder}

The optimized message lengths are applied in the actual LDPC decoder similarly to the W2L mappings, using density evolution.  For each message length $Q_i$ the corresponding syndrome information $I_S$ is determined.  Inverting this mapping, we obtain the function $f^{ds}_Q(I_S)$ that determines the message length $Q$ for the syndrome information $I_S$.  Similarly as before, but now for varying message lengths, the W2L mappings $f^{ds}_{W2L}(w,j)$ are determined.

The two functions $f^{ds}_Q(I_S)$ and $f^{ds}_{W2L}(w,I_S)$ are given to the LDPC decoder in form of pre-computed look-up tables.  After each iteration, the decoder then determines the current value of the syndrome information (see Section~\ref{sec:decoder-state-decoding}), and based on that, it determines the message length and the W2L mapping to be used in the following iteration.

%---------------------------------------------------------------
\section{Simulation Results}
\label{sec:simulation-results}

\begin{figure}[t]
\centering
\psfrag{xlabel}[tc]{${E_{b}}/{N_{0}}$ [dB]}
\psfrag{ylabel}[cb]{bit error rate}
\includegraphics[width=\figwidth]{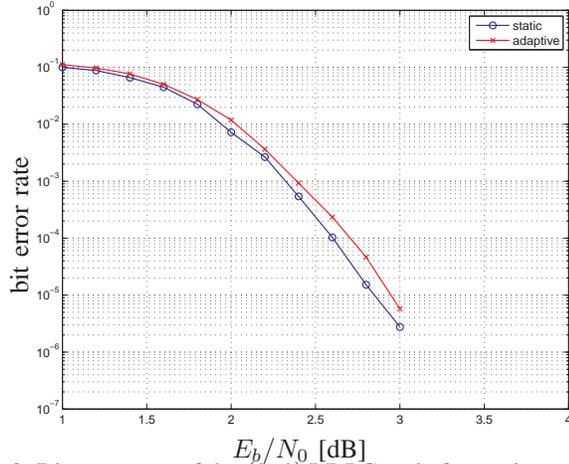}
\caption{Bit error rate of the $(3,6)$ LDPC code for static message lengths $Q=10$ and for dynamic adaptive message lengths $Q \le Q_{\max} = 10$.}
\label{fig:adaptive-ber}
\end{figure}

\begin{figure}[t]
\centering
\psfrag{xlabel}[tc]{${E_{b}}/{N_{0}}$ [dB]}
\psfrag{ylabel}[cb]{sub-iterations}
\includegraphics[width=\figwidth]{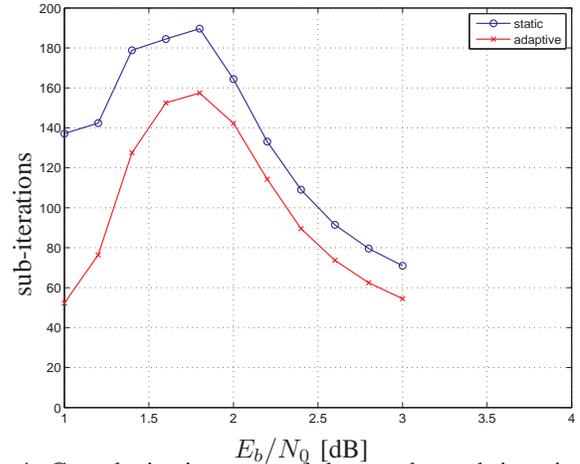}
\caption{Complexity in terms of the number sub-iterations of the $(3,6)$ LDPC code for static message lengths $Q=10$ and for dynamic adaptive message lengths $Q \le Q_{\max} = 10$.}
\label{fig:adaptive-it}
\end{figure}

In this section we demonstrate the performance of the proposed algorithm with a specific example.  We consider a $(3,6)$ LDPC code of length $N=1000$ with $Q_{\max}=10$.  In the following we compare the decoder-state dependent approaches with fixed message length of $Q = Q_{\max}$ and variable message length $Q \le Q_{\max}$.  A quantization of $I_S$ to $20$ levels turned out to be sufficient.

Fig.~\ref{fig:adaptive-ber} shows that the BER for the variable length case is only slightly deteriorated compared to the fixed length case.  The target of the approach is to reduce the average number of sub-iterations, and thus reducing the decoding complexity.  Fig.~\ref{fig:adaptive-it} shows that the average number of sub-iterations of the variable length algorithm is below the fixed length case as expected.

%---------------------------------------------------------------
\section*{Acknowledgment}

This work has been supported in parts by the Australian Research Council under ARC Discovery Grants DP0663567 and DP0558861 and by the ARC Communications Research Network (ACoRN) RN0459498, and by the European Commission in the framework of the FP7 Network of Excellence in Wireless COMmunications NEWCOM++ (contract n. 216715).

%---------------------------------------------------------------
%\IEEEtriggeratref{6}
\bibliographystyle{IEEEtran}
\bibliography{bibnames,isit2009abvmpa}

%---------------------------------------------------------------
\end{document}